\documentclass[final,authoryear,5p]{elsarticle}

\usepackage{graphicx}

\usepackage{amssymb}

\usepackage[
breaklinks=true,%
colorlinks=true,%
pdfauthor={Neilsen},%
pdftitle={Massive, Evolving Winds in Black Hole X-ray Binaries}%
]{hyperref}

\journal{Advances in Space Research}

\begin{document}

\begin{frontmatter}


\title{The Case for Massive, Evolving Winds in Black Hole X-ray Binaries}

\author{Joey Neilsen\corref{cor}\fnref{footnote1}}
\address{Boston University Department of Astronomy, 725 Commonwealth
  Avenue, Room 416D, Boston, MA 02215\vspace{-10mm}}
\cortext[cor]{Corresponding author}
\fntext[footnote1]{Einstein Fellow, Boston University}
\ead{neilsenj@bu.edu}

\begin{abstract}

In the last decade, high-resolution X-ray spectroscopy has revolutionized our understanding of the role of accretion disk winds in black hole X-ray binaries. Here I present a brief review of the state of wind studies in black hole X-ray binaries, focusing on recent arguments that disk winds are not only extremely massive, but also highly variable. I show how new and archival observations at high timing and spectral resolution continue to highlight the intricate links between the inner accretion flow, relativistic jets, and accretion disk winds. Finally, I discuss methods to infer the driving mechanisms of observed disk winds and their implications for connections between mass accretion and ejection processes.
\end{abstract}

\begin{keyword}
accretion, accretion disks; black hole physics; stars: winds, outflows; X-rays: binaries; X-rays: individual (GRO J1655-40)
\end{keyword}

\end{frontmatter}

\parindent=0.5 cm

\section{Introduction}
In the last 20 years, we have seen the discovery of a multitude of
highly-ionized absorbers in moderate and high-resolution X-ray spectra
of black hole and neutron star X-ray binaries
(e.g.\ \citealt{Ebisawa97a,Kotani97a,BS00,K00,Kotani2000a,L02,Sidoli01,Sidoli02,Schulz02,Parmar02,Boirin03,Boirin04,Boirin05,M04a,M06a,M06b,M08,Miller11,NL09,N11a,N12a,N12b,U04,U09,Martocchia06,Kubota07,Blum10,ReynoldsM10,King12,DiazTrigo06,DiazTrigo07,DiazTrigo09,DiazTrigo12a,DiazTrigo12b}). Often
these absorbers are blueshifted, indicative of hot outflowing gas,
i.e.\ accretion disk winds. The prevalence of disk winds in X-ray
binaries suggests that these outflows may play a crucial role in the
physics of accretion and ejection around compact objects. In this
brief review, I discuss some recent developments in the influence
of ionized disk winds around black holes.

\section{Black Hole Accretion Disk Winds and the Disk-Jet Connection}
\label{sec:massive}
Much of the recent work on accretion and ejection processes in black
hole outbursts has focused on radio/X-ray correlations
(e.g.\ \citealt{Gallo03,Corbel03,FB04,FBG04,Fender09}, although see
e.g.\ \citealt{Gallo12} and references therein for lingering questions
about the precise nature of these correlations). Briefly, we now know
that typical black hole transients emerge from quiescence in X-ray
hard states that produce steady, compact jets. They rise in luminosity
in this (probably radiatively inefficient; e.g.\ \citealt{E97})
hard state, until at some point they undergo a transition towards a
much softer state, possibly dominated by a radiatively-efficient
disk. This transition has also been associated with major relativistic
plasma ejections and the disappearance of steady jets. Eventually, the
luminosity falls and they return to quiescence via the hard state. 

Over the last decade, this canonical picture of the ``disk-jet
connection" has proved to be a fruitful way to characterize accretion
and ejection processes around stellar-mass black holes, and has become
the backbone of our understanding of the spectral and timing behavior
of black hole transients. But this story cannot be complete, for it
fails to describe or account for the presence or the influence of
another mode of mass ejection: highly-ionized accretion disk winds,
whose behavior in outburst is only now becoming clear.

Just three years after the launch of \textit{Chandra}, \citet{L02}
argued that winds could be associated with the accretion disk,
although they were not confined to disk-dominated states. \citet{M08}
confirmed that in both GRO J1655--40 and GRS 1915+105, absorption
lines were stronger in spectrally soft states (see also
\citealt{NL09}). They suggested that 
higher ionizing flux might be responsible for the changes in the
winds, but left open the possibility that other (e.g.\ geometric)
changes might be required as well. Thus it remained unclear how or why
winds might change on outburst time scales: were they steady, passive
bystanders that simply responded to variations in the ionizing flux, or
did they play a role in outbursts, appearing and disappearing just like
jets? 

\begin{figure*}
\centerline{\includegraphics*[width=0.45\textwidth]{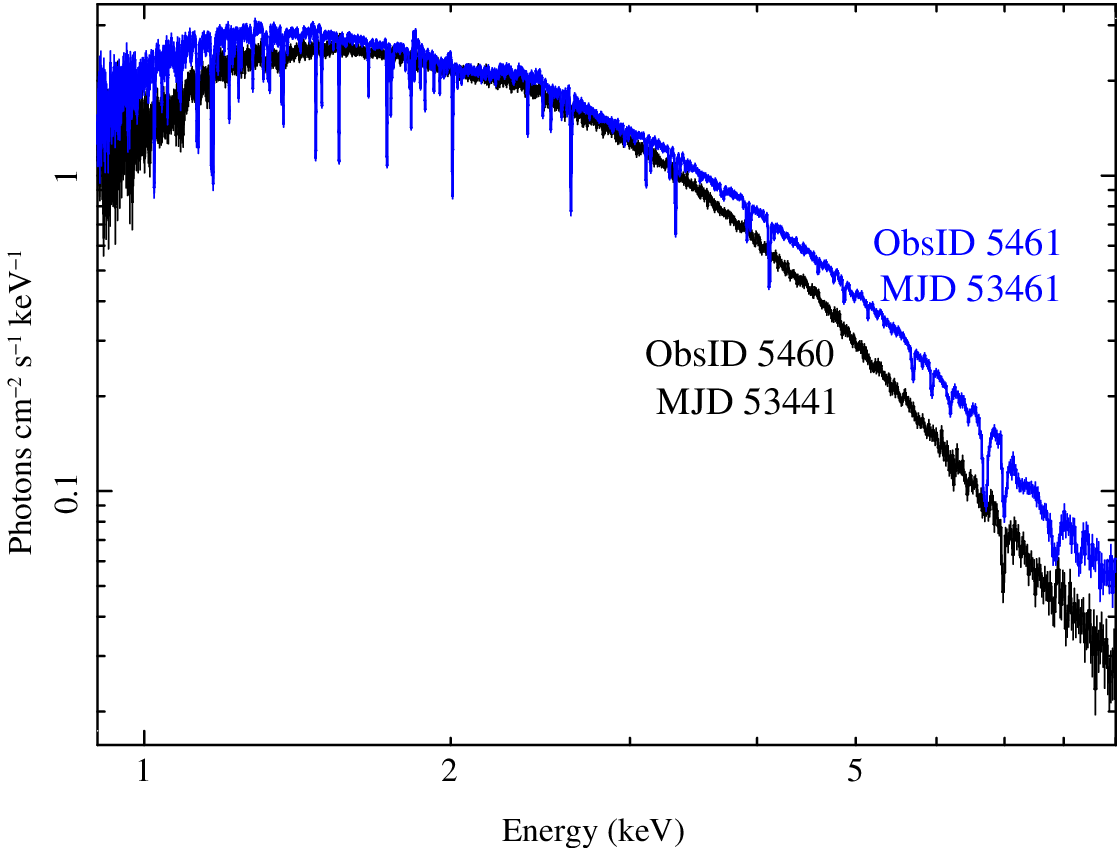}\hspace{5mm}\includegraphics*[width=0.45\textwidth]{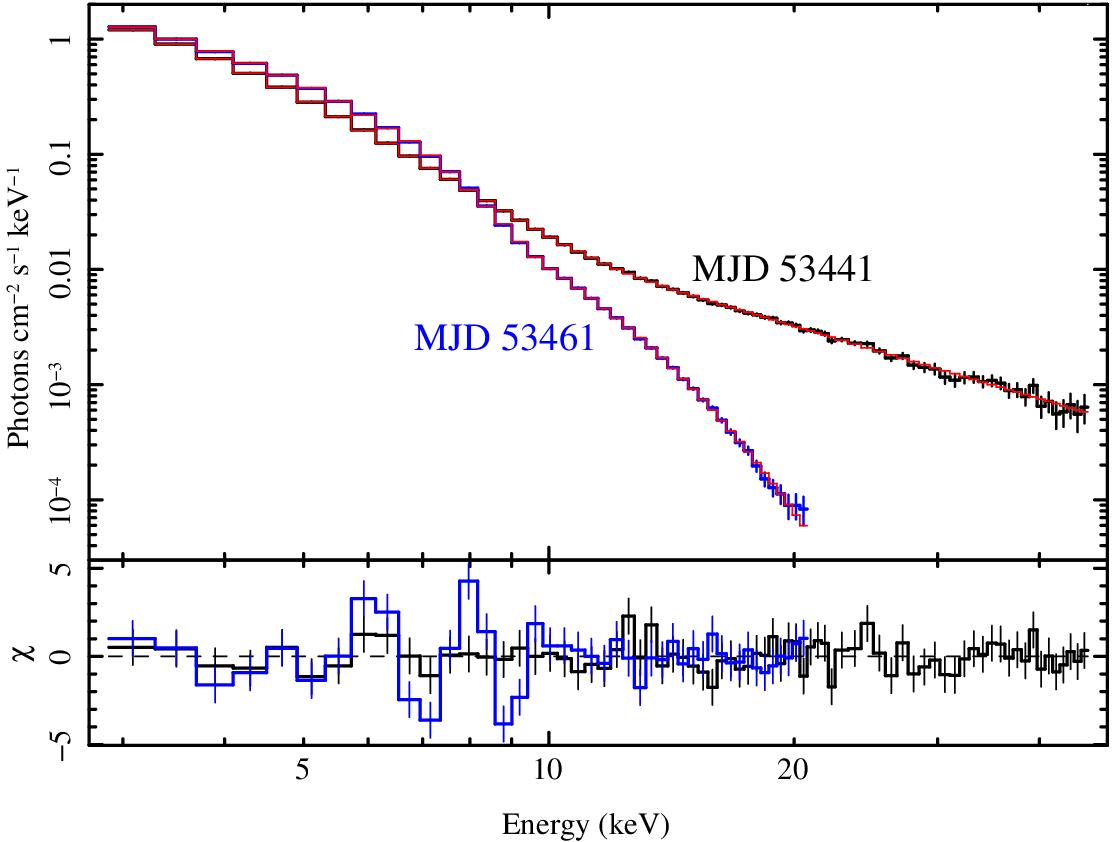}\vspace{-3mm}}
\caption{Spectra of GRO J1655--40 from \citet{N12b}. (\copyright~2012. The American Astronomical Society. All rights reserved.) In both panels, black is the spectrum of the hard state and blue is the spectrum of the softer state. \textbf{Left:} \textit{Chandra} HETGS spectra show only a single line during the first observation, but a rich series of lines from the accretion disk wind in the softer state. \textbf{Right:} \textit{RXTE} PCA show significant differences in the corresponding broadband X-ray spectra, but we argue (\S\ref{sec:case}) that the changes in the ionizing flux cannot explain the differences in the lines.\vspace{-4mm}}
\label{fig:spec}
\end{figure*}
\subsection{A Case Study in Evolving Winds}
\label{sec:case}
With two high-resolution \textit{Chandra} HETGS observations of
accretion disk winds separated by less than three weeks
(\citealt{M08,N12b}), the 2005 outburst of the microquasar GRO
1655--40 presents an ideal backdrop against which to test the
hypothesis that winds do not evolve during outburst. The
\textit{Chandra} and \textit{RXTE} spectra are shown in the left and
right panels of Figure \ref{fig:spec}, respectively. The first
observation (shown in black) took place during a hard state, while the
second observation (shown in blue) occurred during a much softer
state. And while the first observation contained an Fe\,{\sc xxvi}
absorption line near 7 keV, the second provided an extremely rich
absorption line spectrum that has been studied in great detail
(\citealt{M06a,Netzer06,M08,Kallman09,N12b}; see \S\ref{sec:driving} for a
discussion of the origin of this wind). 

\begin{figure}
\centerline{\includegraphics*[width=3.2in]{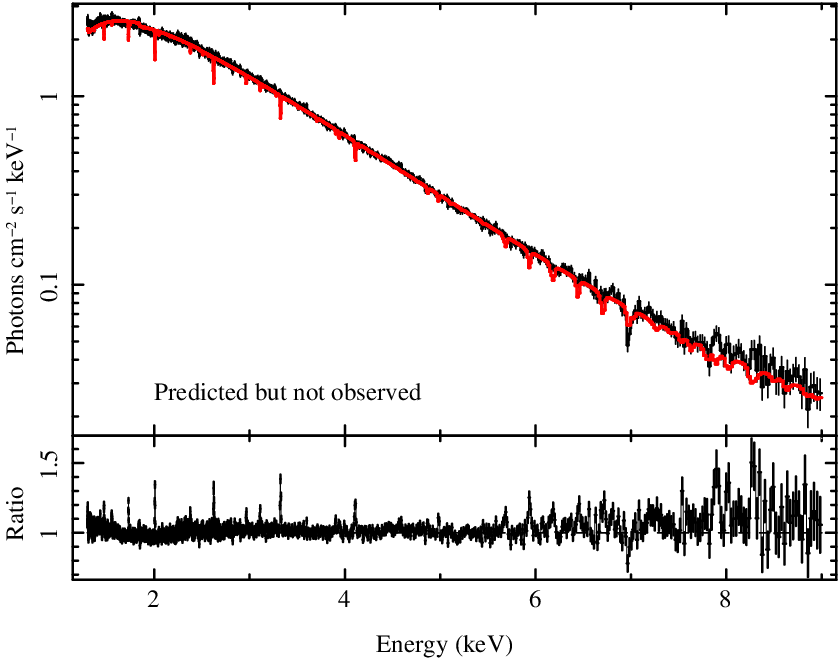}}\vspace{-3mm}
\caption{Photoionization models of a steady wind in GRO J1655--40 from \citet{N12b}. (\copyright~2012. The American Astronomical Society. All rights reserved.) Based on \citet{Kallman09}. If the same wind were present in both \textit{Chandra} HETGS observations, we should have detected a number of strong absorption lines during the hard state. The absence of these lines indicates wind variability over the course of the outburst.\vspace{-5mm}}
\label{fig:xstar}
\end{figure}

Here, let us consider the question: why are the two \textit{Chandra}
absorption line spectra so different? Are the differences driven by
changes in the photoionizing flux from the hard state to the soft
state, or did the wind physically evolve over those 20 days? Our
detailed analysis (\citealt{N12b}) indicates that the wind must have
evolved significantly between the two \textit{Chandra} observations. 
This argument can be understood both qualitiatively and quantitatively: 
\begin{enumerate}
\item A comparison of the hard state and soft state PCA spectra in
  Figure \ref{fig:spec} reveals a clear excess of photons with $E>10$
  keV, which we usually think of as ionizing photons. Thus, at first
  glance it seems plausible that changes in the ionizing flux could
  explain the differences in the lines. In fact, however, \textit{the
  ionization of this wind is determined primarily by soft X-rays},
  since many of the visible ions during the softer state, like O, Ne,
  Na, Mg, Al, and Si, are effectively transparent to hard X-rays (due
  to their small cross-sections above 10 keV). Since the soft X-ray
  spectra of the two observations are quite similar, we conclude that
  the change in the relevant ionizing flux is negligible and cannot,
  in and of itself, explain the observed differences in the
  lines.
  \item The physical properties of the rich absorber during the soft
  state are well known (\citealt{M06a,M08,Kallman09}), so we can use
  photoionization codes like {\sc xstar} (\citealt{Bautista01}) and
  the 1 eV -- 1 MeV radiation field to generate predictions about
  ionized absorption during the hard state. Our results (Figure
  \ref{fig:xstar}; \citealt{N12b}) clearly indicate that a number of
  strong absorption lines \textit{would have been} visible during the
  hard state if the wind had been steady; the non-detection of these
  lines confirms that the wind must have evolved during those 20
  days. For several simple but realistic scenarios for the geometrical
  evolution of the wind (\citealt{N12b}), we argue that the variations
  in its ionization and column density likely imply an increase in the
  density and mass loss rate in the wind by a factor between 25 and
  300.
\end{enumerate}
To summarize briefly, after \citet{L02}: ``ionizing flux is only part
of the solution." Based on our careful treatment of photoionization,
we find compelling qualitative and quantitative evidence for
significant physical changes in the accretion disk wind during the
2005 outburst of GRO J1655--40. In the following section, we argue
that broad parallels between this source and other black holes support
the conclusion that evolving winds may be an extremely common, if not
universal, phenomenon. 

\subsection{Ubiquitous, Massive Evolving Winds}
As noted above, \citet{L02} and \citet{M08} pointed out that accretion
disk winds seem to be associated with the accretion disk and/or
spectrally soft states. Recent work by \citet{Ponti12} provides
convincing evidence for these earlier claims: their archival study of
\textit{Chandra} HETGS, \textit{XMM-Newton}, and \textit{Suzaku}
observations of stellar mass black holes in outburst demonstrates that
accretion disk winds are preferentially detected in softer
states\footnote{Note that disk winds are only detected in systems seen
  at high orbital inclination, which is consistent with the
  interpretation that these absorbers are localized near the
  equatorial plane of the disk. Based on the inclinations of systems
  with detected disk winds, \citeauthor{Ponti12} estimate a typical
  wind opening angle $\theta\lesssim30^{\circ}$. This conclusion
  echoes established results from studies of neutron star
  low-mass X-ray binaries, where X-ray absorbers are known to be
  concentrated near the disk plane
  (e.g.\ \citealt{Sidoli01,Boirin03,DiazTrigo06}).}. In particular,
winds are ubiquitous along the high-luminosity branch of the
hardness-intensity diagram after the spectrally hard state (see
\citealt{L02,M08,N12b} for rare cases of weak winds at the high 
luminosity end of spectrally hard states). In some cases, stringent upper limits have been placed on the existence of hard state winds (e.g.\ \citealt{Blum10,Miller12}). 

\citet{Ponti12} suggest that a static absorber with a variable 
ionization parameter may be unlikely to explain completely the
observed behavior of winds in black hole outbursts, although it is
also noted that ionization effects may be important and that only
detailed photoionization studies can confirm this suggestion. While
this is certainly true, black hole wind variability studies on time
scales from seconds (\citealt{N11a}) to hundreds or thousands of
seconds (\citealt{L02,M06b}) to weeks and years
(\citealt{N12b,Blum10,Miller12}) have all required changes in the wind
density. It therefore seems likely that the observed outburst behavior
of winds will also require such changes, in which case we can conclude
that disk winds are preferentially but not exclusively
\textit{launched} at high luminosity, around or after the time
the black hole begins to exit the spectrally-hard state. 

\begin{figure*}
\centerline{\includegraphics*[width=0.45\textwidth]{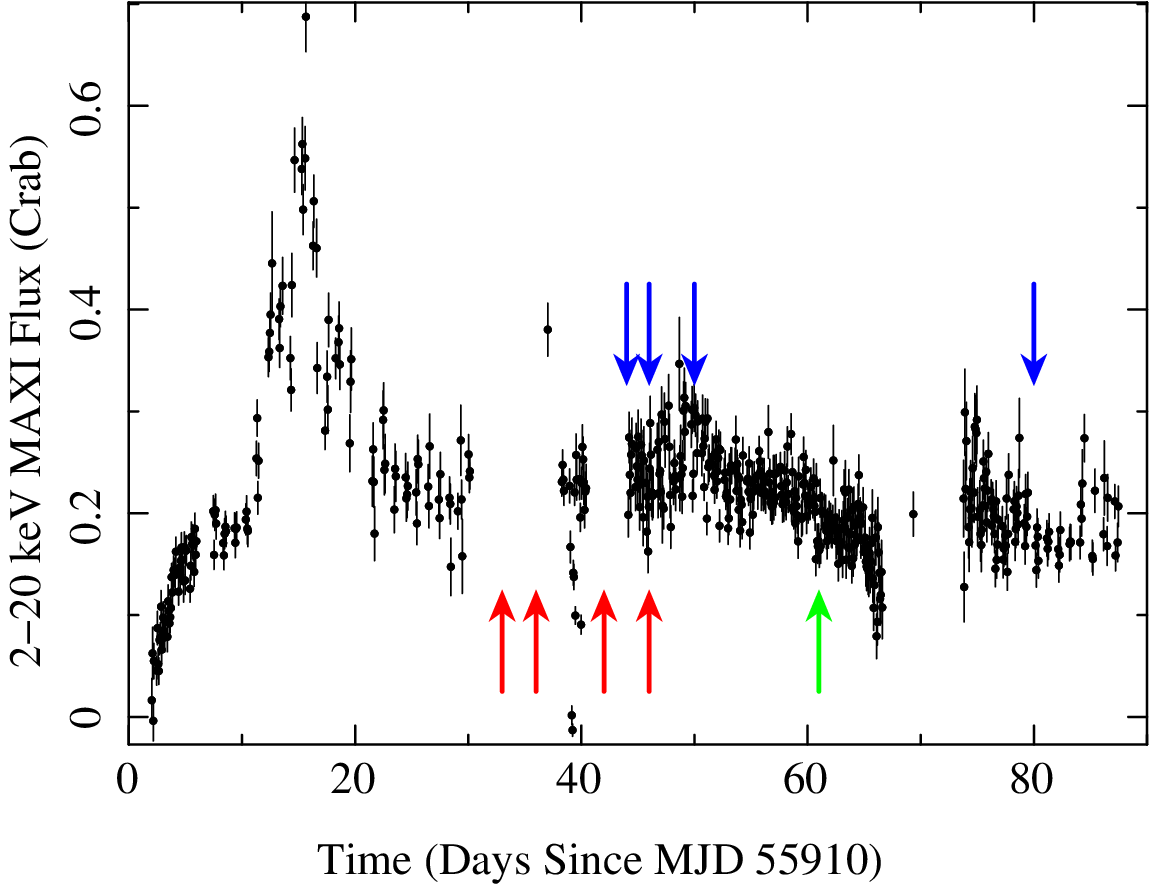}\hspace{5mm}\includegraphics*[width=0.45\textwidth]{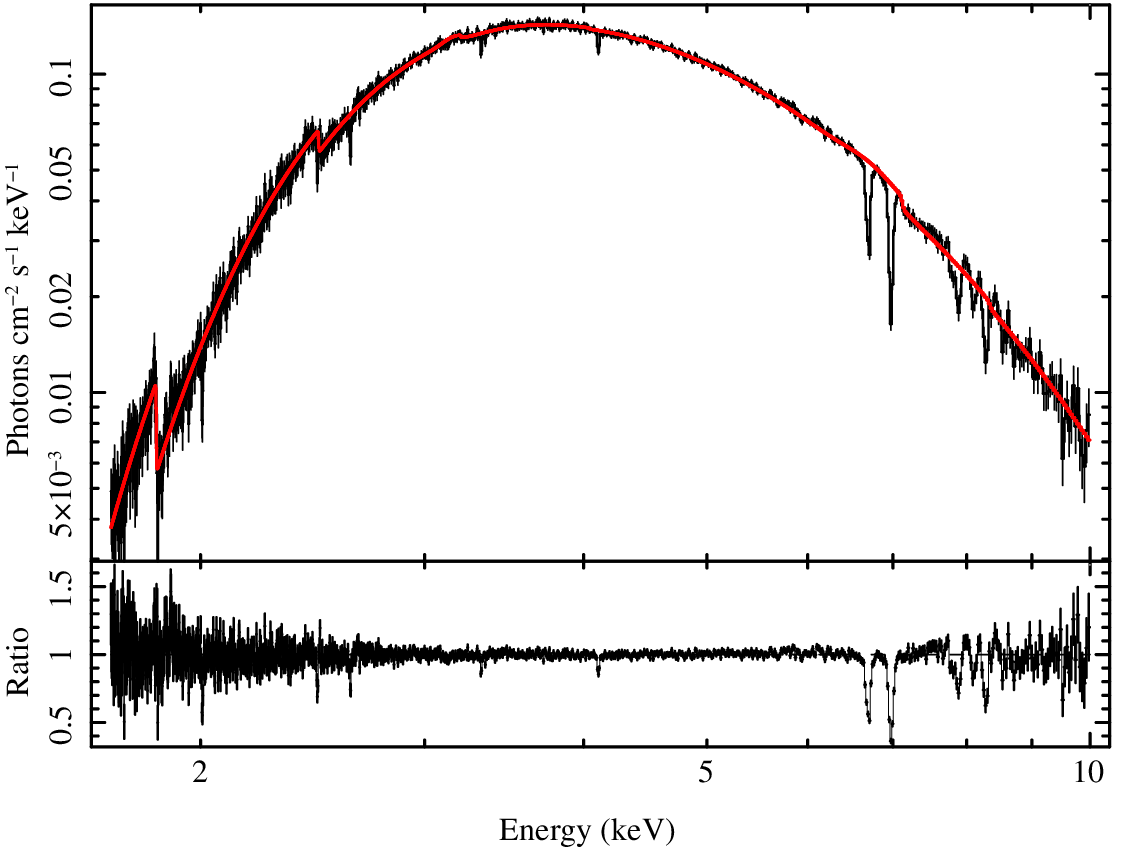}}
\caption{\textbf{Left:} \textit{MAXI} monitoring of the 2011-2012
  outburst of 4U 1630--47, with \textit{Chandra} HETGS and
  \textit{Suzaku} observations indicated by upward arrows, and other
  wavelengths indicated by downward arrows. \textbf{Right:} 
  \textit{Chandra}'s high-resolution spectra reveal a strong accretion
  disk wind at precisely the phase of the outburst indicated by
  \citet{Ponti12}.}  
\label{fig:1630}
\end{figure*}

To test this interpretation, we undertook a \textit{Chandra} HETGS
campaign to catch this phase of a new outburst; the resulting spectra 
of 4U 1630--47 are shown in Figure \ref{fig:1630}. The results will be
published in detail in future work (\citealt{N12e}), but suffice it to
say here that with the robust detection of a strong outflow, this
campaign was remarkably successful. It should be noted that
\citet{Kubota07} detected a wind in a similar phase of a prior
outburst of 4U 1630--47, so our new detection confirms that wind
behavior is predictable. We conclude that winds are reliably launched
during this outburst phase in black hole X-ray binaries; confirming
their absence or weakness in harder and less luminous states will be
the subject of future work.

If winds were simply ionized gas along the line of sight, such a
conclusion might be interesting but relatively insignificant. In
reality, there is now a large and growing body of evidence indicating
that disk winds in stellar mass black holes may be extremely
massive. In fact, as early as a decade ago, it was discovered that
wind mass loss rates $\dot{M}_{\rm w}$ could be comparable to black
hole accretion rates $\dot{M}_{\rm acc}$ \citep{L02}. More recently
\citet{NL09} suggested that radiatively/thermally-driven winds could
deplete the mass of the disk enough to suppress relativistic jets. In
a few exceptional cases (e.g.\ the `heartbeat' state of GRS 1915+105,
\citealt{N11a};  IGR J17091--3624, \citealt{King12}), detailed studies
have found mass loss rates in excess of $(10-20)~M_{\rm acc}$! These
remarkable results, too, are supported by the results of
\citet{Ponti12}, who find that $\dot{M}_{\rm w}$ is typically at least
twice $\dot{M}_{\rm acc}$, and approaches $10\dot{M}_{\rm acc}$ at
high Eddington ratio. 

If winds are truly as massive as these results suggest, it begins to
seem significant that they are preferentially launched at the same
phase of black hole outbursts when we observe the disappearance of
steady jets and major changes in the structure of the accretion
flow. Could it be that disk winds are indeed the mechanism by which
jets are suppressed and state transitions take place, as suggested by
\citet{NL09} and \citet{N11a}? At present, the data cannot rule out
this interpretation, but with careful tracking of $\dot{M}_{\rm w}$
going into this state transition, it may be possible to shed new light
on this important question in the near future.

\section{On Inferring Wind Driving Mechanisms}
\label{sec:driving}
As noted in \S\ref{sec:massive}, in the last few years there have been
a number of developments that suggest deep connections between the
X-ray luminosity or accretion rate, the state of the accretion flow,
and the behavior of accretion disk winds. The significance of such
connections is not entirely clear, however: \citet{Miller12} (see also
\citealt{M08}) have argued that these some of these connections can be
explained in terms of a magnetic field configuration that changes
during outburst, while other authors have presented interpretations
based on radiatively- and thermally-driven winds
(\citealt{NL09,U09,U10,N11a,N12a,Ponti12}). 

Because the detailed implications of these massive evolving winds
depend heavily on their formation physics, it is critical to draw
robust conclusions about the mechanisms that produce them. To this
end, we often take advantage of the fact that the well-known driving
mechanisms (radiation pressure, Compton heating, and MHD) typically
operate in different regimes of density, ionization, and distance from
the black hole (e.g. \citealt{PK02,M08}). 

For example, since radiation pressure is most commonly transmitted via
UV resonance absorption lines, it may be ineffective when the gas has
little or no opacity in the UV (e.g.\ at very high ionization,
$\xi\gtrsim10^{3}$ ergs cm s$^{-1}$; \citealt{PK02}). In contrast,
Compton heating may produce highly-ionized outflows, but because they
require a large surface area of gas to be heated to the point that the
sound speed exceeds the escape speed, thermally-driven winds are only
expected at large distances from the black hole ($10^4-10^5$ $r_g$;
\citealt{B83,Woods96}). In addition, simulations at varying
luminosities and spectral shapes (e.g.\ \citealt{Woods96})
have shown that thermal driving tends to produce outflows with small
mass fluxes and/or gas densities ($n\lesssim10^{12}$ cm$^{-3}$ for the
specific case of GRO J1655--40; \citealt{Luketic10}). MHD 
processes like the Blandford-Payne mechanism, however, may produce
dense outflows at small radii \citep{BlandfordPayne}. 

\subsection{Absorption Lines, Density, and GRO J1655--40}
In principle, then, it is possible to use the observed properties of
winds to infer their launching mechanisms. The difficulty is that many
physical factors influence the observability of lines. For example,
the X-ray luminosity sets the ionization parameter of the
gas: \begin{equation}\xi=\frac{L}{n r^2},\end{equation} while the
shape of the ionizing spectrum and the gas density determine which
ions are visible at any $\xi$ (\citealt{Kallman01}). The density and
geometry of the wind determine the equivalent hydrogen column density
of the absorber: \begin{equation}N_{\rm H}=n\Delta r,\end{equation}
and the column density of each individual ion follows from $N_{\rm
  H},$ the chemical abundances $A_i,$ and the ionization balance
$x_i$: \begin{equation}N_i=x_iA_iN_{\rm H}.\end{equation} Finally, the
ion column densities are folded into the equivalent width
$W_{\lambda}$ of each line via the curve of growth\footnote{Note that
this equation only holds in the optically-thin limit.}: 
\begin{equation}\frac{W_\lambda}{\lambda}=\frac{\pi e^2}{m_e c^{2}}N_i
\lambda f_{ji}. \end{equation} Here $\lambda$ and $f_{ji}$ are the
line wavelength and oscillator strength, respectively. Given the
complexity of the connections between the gas properties, the
radiation field, and illumination patterns, it is not immediately
obvious how any observed wind is produced. However, for a wind that is
both very highly ionized ($\xi\gtrsim10^{3}$ ergs cm s$^{-1}$) and
sufficiently dense that its implied radius (c.f.\ Equation 1) is well
inside the radius where thermal driving is effective (i.e.\ the
Compton radius, $R<<R_{\rm C}\sim10^{11-12}$ cm), the natural conclusion
is that MHD processes likely play a role in its launching.  

The classic example of this argument in black hole X-ray binaries
comes from \citet{M06a}, who first published the extraordinary
\textit{Chandra} HETGS absorption line spectrum of GRO J1655--40
(Figure \ref{fig:spec}). Photozioniation analysis indicated a
characteristic ionization parameter of $\xi\gtrsim10^4$ ergs cm
s$^{-1},$ an order of magnitude too high for line driving to be
effective. In addition to many other lines, they detected two Fe\,{\sc
xxii} absorption lines at 11.77 \AA~and 11.92 \AA~whose ratio can be
used as a density diagnostic. Their analysis and subsequent studies
(\citealt{M08,Kallman09}) led to the conclusion that the density must
have been at least $n\gtrsim 10^{14}$ cm$^{-3},$ placing the absorber
some three orders of magnitude inside the Compton radius, where
thermal driving cannot operate. Based on the high optical depth in the
wind and possible saturated lines, \citet{Netzer06} argued for a lower
density ($n_{e}\sim10^{13}$ cm$^{-3}$) and ionization parameter
($\xi\sim10^{3}$), which would have implied a more distant wind
consistent with thermal driving. \citet{M08} subsequently argued that 
\citeauthor{Netzer06}'s model significantly overpredicted  
soft X-ray absorption lines, and according to simulations of
thermally-driven winds tailored to GRO J1655--40
(\citealt{Luketic10}), even the lower density proposed by 
\citet{Netzer06} is an order of magnitude too high for thermal driving
to be the dominant launching mechanism. By process of elimination,
\citet{M06a,M08,Kallman09} concluded that the dense wind in GRO
J1655--40 must be powered by magnetic processes\footnote{This was a
  long-awaited discovery, particularly since it had been known for
  some time that both accretion disks and relativistic jets are
  governed in part by magnetic processes
  (e.g.\ \citealt{Balbus91,BlandfordZnajek,McKinney06,Mirabel99} and
  references therein).}. 

\subsection{MHD Winds in Black Hole X-ray Binaries?}
But how are we to understand the disk-wind-jet coupling in light of
the apparent variations in wind formation physics between different
systems? Radiative/thermal driving has been invoked to explain most
winds in black hole systems, but there is convincing evidence for an
MHD wind in GRO J1655--40. The continuing debate over the accretion
disk wind launching mechanism begs the question: \textit{is there} a
single, universal process that governs the interaction between disks,
winds, and jets in black hole X-ray binaries? Can the behavior of
inflows and outflows be unified in such disparate systems? As our
understanding of accretion and ejection physics evolves, it can be
instructive to revisit prior observations and the conclusions we draw
from them.

\defcitealias{Reynolds12}{R12}
Perhaps the most salient development comes from \citet{Reynolds12},
hereafter \citetalias{Reynolds12}, who demonstrates that the
well-known launching mechanisms (i.e.\ Compton heating,
magnetocentrifugal acceleration, and radiation pressure) can be
distinguished not only by the ionization and the density, but also the
optical depth of the winds they produce. With a focus on Compton-thick
winds, the forbidden regime of parameter space for each mechanism is
clearly set out in terms of the optical depth $\tau,$ the Eddington
ratio $\lambda,$ the ratio $f_v$ of the wind's terminal velocity to
the escape velocity at the launch radius, and the distance to the
black hole $r$ (in units of gravitational radii $r_g$). 

How does this analysis inform the interpretation of the dense,
highly-ionized wind in GRO J1655--40, which had a column density
$N_{\rm H}\sim10^{24}$ cm$^{-2}$ (near the Compton-thick limit;
\citealt{M06a,M08,Kallman09})? Following \citetalias{Reynolds12}, let
us consider each launching mechanism in turn: 

\begin{enumerate}
\item \textit{Thermal Driving}: As noted above, and by
  \citet{M06a,M08}, and \citet{Luketic10}, thermal driving cannot
  possibly produce such a dense outflow close to the black hole
  ($r\sim(1-7)\times10^9$ cm $=1000-6800~r_g$;
  \citealt{Kallman09}).\vspace{2mm} 
\item \textit{Radiative Driving}: Again, as argued above following the
  original work on these data, the wind is far too ionized for UV line
  driving to be effective at launching the wind. But radiation
  pressure can also act on free electrons, and \citetalias{Reynolds12}
  shows that in this case, the momentum transferred to the electrons
  is insufficient to drive a wind if $\lambda<2f^2_v.$ In other words,
  the momentum flux in a radiatively-driven wind cannot exceed that of
  the radiation field.

\hspace{3mm} The soft-state wind of GRO J1655--40 provides an
interesting illustration of this constraint. During this particular
observation, we estimate the Eddington ratio to be about $\lambda\approx0.06$
(\citealt{N12b}). The ratio $f_v$ is much harder to determine, since
only the line-of-sight velocity can be measured. \citet{Kallman09}
found a blueshift of $\sim375$ km s$^{-1},$ which is much less than
the escape velocity at plausible launch radii ($v_{\rm
  esc}\sim5000-14,000$ km s$^{-1}$; $f_v \sim0.03-0.07$). Normally,
one might suppose that $f_v\sim 1$ and attribute the small blueshift
to a velocity primarily perpendicular to the line of sight, but it is
difficult to see how this scenario is consistent with the small solid
angle of the wind (see below). 

\hspace{3mm} If we take the velocities at face value and accept that
the wind remains bound to the black hole, we find
$\lambda\sim(6-40)\times2f^2_v,$ i.e.\ there is no shortage of
momentum in the radiation field. However, this is not a sufficient
condition to launch the wind, since the radiation force on the gas is
still required to exceed the force of gravity (C.\ Reynolds 2012, private
communication). Thus, despite an abundance of momentum flux, radiation 
pressure cannot explain the dense, highly-ionized wind in GRO
J1655--40 (see also \citealt{M06a}).\vspace{2mm} 
\item \textit{Magnetocentrifugal Driving}: Although magnetocentrifugal
acceleration (e.g.\ \citealt{BlandfordPayne}) is not the only MHD
process that can drive winds (e.g.\ \citealt{Proga03}; see also
\citealt{M08} and references therein), it has been studied in great
detail and lends itself nicely to an analytic constraint on the
production of Compton-thick outflows. As shown by
\citetalias{Reynolds12}, Compton-thick magnetocentrifugal winds can
only be produced at radii \begin{equation} 
r<800~\varpi^{-7}\left(\frac{\tau}{\lambda}\right)^{-2}\left(\frac{\Omega}{\pi}\right)^{-2}\left(\frac{\eta}{0.1}\right)^{-2}r_g,
\end{equation}
i.e.\ where 
\begin{equation}
\frac{\tau}{\lambda}<\sqrt{\frac{800~r_g}{\varpi^7 r}}\left(\frac{\Omega}{\pi}\right)\left(\frac{\eta}{0.1}\right).
\end{equation} Here $\Omega$ is the solid angle of the wind, $\eta$ is
the radiative efficiency of the accretion flow, and $\varpi\sim2-3$ is
the ratio of the size of the acceleration zone of the wind to the
launch radius (for more details, see \citetalias{Reynolds12} and
references therein).  

\hspace{3mm} For the dense wind in GRO J1655--40, the observed column
density $N_{\rm H}=10^{24\pm0.02}$ cm$^{-2}$ implies an optical
depth\footnote{Since the vast majority of the electrons in an ionized
plasma come from hydrogen, for our purposes it is reasonable to
assume an ionization fraction very close to 1.} $\tau\sim0.67$, from
which we estimate $\tau/\lambda\sim11\pm2.$ \citet{M06a} use upper
limits on emission line strengths to place the tight constraint
$\Omega<4\pi/9.$ If we allow $\varpi\gtrsim1$ and use the smallest
plausible radius $r\sim970~r_g,$ Equation 6 for the allowed parameter
space for magnetocentrifugal winds becomes $\tau/\lambda<2.0.$ That
is, given the small solid angle of the wind and its relatively large
distance from the black hole, the optical depth of the wind is at
least $\sim5\times$ too large for it to be launched by
magnetocentrifugal processes. However, we stress that other MHD
mechanisms (see \citealt{M08} and references therein) have not been
ruled out. 
\end{enumerate}
In short, based on the constraints presented in \citet{Reynolds12}, we
find that the dense, highly-ionized wind in GRO J1655--40 cannot be
driven by magnetocentrifugal effects. We therefore confirm the
original suggestion of \citet{M08} that the wind is driven by some
other MHD process, like magnetic pressure
(e.g.\ \citealt{Proga03}). It should be noted that
\citetalias{Reynolds12} uses a simplified model of the wind and
driving mechanisms, and that the geometry of observed winds may be
somewhat more complicated. For example, as detailed by
\citet{Giustini12} and references therein, the bulk properties of 
winds (e.g.\ density, ionization, velocity, etc.) may be strong,
non-monotonic functions of position, even for outflows with simple
streamlines. This is clearly an area where future theoretical work can
continue to improve the accuracy and robustness of inferences from
observations. In any case, more detailed study of the remarkable wind  
in GRO J1655--40 is forthcoming.

Finally, we can return to the question at hand: is there a single,
universal process that governs the interaction between disks, winds,
and jets in black hole X-ray binaries? To the best of our knowledge,
we can trace the origin of most winds from stellar-mass black holes
back to the radiation field of the inner accretion flow
(e.g.\ \citealt{L02,Kubota07,NL09,U09,U10,N11a,N12a}). This seems to 
be a compelling argument for the scenario presented in
\S\ref{sec:massive}, in which accretion disk winds play an integral
role in the evolution of black hole outbursts \textit{by virtue of}
their connections to the radiation field. 

However, given that there is one clear case of a magnetically-driven
wind, and no hard evidence that other winds are \textit{not} driven by
magnetic fields, this question should still give us pause. Are the
results from detailed studies of one wind observed in GRO J1655--40
applicable to the class of black hole binaries as a whole?
\citet{King12b} find evidence of a three-way correlation between jet
power, wind power, and bolometric luminosity over eight orders of
magnitude in black hole mass. This seems to  indicate that outflows
and inflows could be regulated by a common process related to the mass
accretion rate, but the underlying physics of this regulation is still
open for discovery.  

\section{Conclusions}
In the last few years, a significant effort has been devoted to
understanding the physics and behavior of accretion disk winds in
black hole X-ray binaries, with important developments coming from
both archival studies and new observations. By extrapolating from
the specific (exceptional) case of GRO J1655--40 to the ensemble of
winds studied by \citet{Ponti12}, we have come to understand that
highly-ionized winds are ubiquitous around stellar mass black holes,
that they may evolve significantly in outburst, and that they may
carry away a significant fraction of the inflowing gas. It is worth
reiterating that winds are not necessarily confined to specific states
\textit{per se}, but appear to evolve \textit{continuously} during
outbursts. While their formation physics is complex and challenging to
discern, and while MHD processes may be important, it seems that most
(but not all) known accretion disk winds are consistent with radiative
or thermal driving (\citealt{L02,Kubota07,NL09,U09,U10,N11a,N12a};
\citealt{DiazTrigo12b} come to a similar conclusion for neutron star
binaries). 

At once driven by and ionized by the luminosity
of the central engine, these massive outflows may require radiation for
their existence, but their substantial influence may ultimately be
their undoing. By draining vast quantities of mass from the accretion
disk, they may not only suppress relativistic jets and cause or
facilitate state transitions as seen in GRS 1915+105 (with tentative
evidence in other systems; \citealt{Ponti12}), but they may also
cripple the ability of the disk to launch a wind! In the future, by
taking advantage of our ever-growing understanding of the
spectral/timing behavior of X-ray binaries in outburst, we will track
these evolving, massive, ionized outflows in order to continue
shedding new light on the physics of accretion and ejection around
black holes. 

\section*{Acknowledgements} I thank the anonymous referees, whose comments enhanced the context and clarity of the paper, as well as Chris Reynolds and Jon Miller for
comments that substantially improved the discussion of wind
driving mechanisms. This work was supported by the National
Aeronautics and Space Administration through the Smithsonian
Astrophysical Observatory contract SV3-73016 to MIT for support of the
\textit{Chandra} X-ray center, which is operated by the Smithsonian
Astrophysical Observatory for and on behalf of the National
Aeronautics and Space Administration under contract NAS8-03060, and by
NASA through the Einstein Fellowship Program, grant PF2-130097.

\bibliographystyle{model5-names}
\bibliography{jasr}

\end{document}